\documentclass[english,twocolumn, prl,nofootinbib,nobibnote,showpacs]{revtex4}
\usepackage[T1]{fontenc}
\usepackage{amstext}
\usepackage{amsmath}
\usepackage{amssymb}
\usepackage{graphicx}
\usepackage{esint}
\usepackage{units}

\makeatletter
\@ifundefined{textcolor}{}
{%
 \definecolor{BLACK}{gray}{0}
 \definecolor{WHITE}{gray}{1}
 \definecolor{RED}{rgb}{1,0,0}
 \definecolor{GREEN}{rgb}{0,1,0}
 \definecolor{BLUE}{rgb}{0,0,1}
 \definecolor{CYAN}{cmyk}{1,0,0,0}
 \definecolor{MAGENTA}{cmyk}{0,1,0,0}
 \definecolor{YELLOW}{cmyk}{0,0,1,0}
}

\usepackage{graphicx,color,times}
\usepackage{eucal}
\usepackage{hyperref}
\definecolor{darkred}{rgb}{0.8,0.1,0.1}
\hypersetup{colorlinks=true, linkcolor=darkred, citecolor=blue}

\makeatother

\usepackage{babel}
\begin{document}

\title{Edge exponents in work statistics out of equilibrium and dynamical phase
transitions from\\ scattering theory in one dimensional gapped systems}

\author{T. Palmai}

\email{palmai@phy.bme.hu}

\affiliation{MTA-BME ``Lendulet'' Statistical Field Theory Research Group, Institute of Physics, Budapest University of Technology and Economics, Budafoki ut 8, 1111 Budapest, Hungary}
\begin{abstract}
I discuss the relationship between edge exponents in the statistics
of work done, dynamical phase transitions, and the role of different
kinds of excitations appearing when a non-equilibrium protocol is
performed on a closed, gapped, one-dimensional system. I show that the edge 
exponent in the probability density function of the work is insensitive to 
the presence of interactions and can take only one of three values: $+1/2$, $-1/2$ 
and $-3/2$. It also turns out that there is an interesting interplay between 
spontaneous symmetry breaking or the presence of bound states and the 
exponents. For instantaneous global protocols, I find that the presence
of the one-particle channel creates dynamical phase transitions in the time 
evolution.
\end{abstract}

\maketitle
Out of equilibrium phenomena in quantum systems have been given a
large amount of attention recently. The interest was largely spun
by the advent of new experimental techniques in cold atoms 
and solid state quantum devices where coherence can
be maintained for far longer times than previously \cite{coldat}, and therefore
the unitary evolution after a quantum system is taken out of equilibrium
has become an important and well studied concept. This has been renewing 
interest
in some fundamental and long-standing questions in statistical mechanics,
and at the same time bringing new ideas and phenomena into the spotlight.
One such concept is that of dynamical phase transitions (DPTs), which refers
to nonanalytical behavior detected in the Loschmidt echo (LE) \cite{Heyl} and affecting the time evolution of certain observables in a characteristic way \cite{Heyl2}. For
the important class of global, instantaneous, nonequilibrium protocols
(dubbed as quantum quenches), this phenomenon can be understood in
terms of the Fisher zeros of the partition function corresponding
to singularities of the free energy: The LE in this case
is equivalent to the partition function with imaginary temperature
\cite{Heyl}. While DPTs have been the subject of a growing number
of both analytic and numerical works, a clear physical mechanism
accounting for them has yet to emerge \cite{Heyl,Heyl2,SilvaSmacch,AS,Dora,newrefs}.

Another interesting quantity is the work performed when taking the
system out of equilibrium \cite{Silva2008}. With the discovery of 
nonequilibrium
fluctuation relations \cite{Jarz} this is interesting on its own
right, but it is also intimately connected to the LE for
certain important protocols \cite{Talkner}: In the case of quantum quenches,
the LE and the probability density function (PDF) of the work done
are related by Fourier transformations. Furthermore,
it seems now that although the work itself is not an observable 
\cite{Talkner}, due to being a positive operator valued measure, it can in 
principle be measured
on an enlarged system \cite{Roncaglia}. One
of the most striking features of the statistics of work is the robustness
and universality of the edge singularity exponent in its PDF at the lower
limit, corresponding to the opening
of the first continuous channel of realizing the quench, i.e., the emission of
two (quasi-)particles with opposite momenta \cite{SilvaSmacch,GS,PalmaiSot}. 
This robustness has already been demonstrated with respect to the details of the 
protocol \cite{SilvaSmacch}.

In this paper we will concentrate on the role of interactions, and we will 
determine the possible exponents emerging from
the statistics of work in one-dimensional gapped systems. We will connect
the different values to different kinds of quasiparticle contents.
We establish that the crucial property is the existence or absence
of one-particle excitations, which can appear, e.g., in the form of bound states 
or
when the initial or the final system is spontaneous symmetry breaking. 
We also find that the exponent is extremely robust
and, in fact, close to criticality there are only three possible values
(excluding fine tuning): $+1/2$, $-1/2$ and $-3/2$ independent 
of the relevant critical point and the symmetries of the system. Our results are 
also interesting
 with respect to DPTs: For global quenches we can predict
the emergence of a transition by looking at the pre- and postquench
particle contents. 

In the following, we first discuss the possible edge singularity
exponents through a scattering theoretical argument. We then study
the case of spontaneous symmetry breaking on the example of the Ising
model. Then we move on to discuss the sine-Gordon model, which
provides a low-energy effective field theory description of many interesting 
condensed matter systems, e.g., one-dimensional magnets of the $XYZ$ and $XXZ$ types 
and Mott insulators \cite{KonikEssler}. Finally, the connection to the LE is 
studied.

\paragraph{Edge exponent from scattering theory.}

We apply quantum field theoretical scattering theory to extract the
exponents. This approach is natural since the edge exponent is determined
only by the low-energy part of the spectrum, and quantum field theory gives
the universal low-energy effective description valid close to criticality.

Suppose we perform some finite $T$ time nonequilibrium protocol on our system
\begin{equation}
H[g(t_0)]=H_{0}\rightsquigarrow H_{1}=H[g(t_0+T)],
\end{equation}
beginning in, e.g., the ground state of an initial Hamiltonian $H_{0}$,
which is allowed to evolve by a different, local Hamiltonian that
may itself be explicitly time dependent through a
coupling, e.g., the magnetic field $H[g(t)]$. 
At the end of the
protocol we arrive in some state that can be expanded in terms
of asymptotic states of the final Hamiltonian $H_{1}$. Asymptotic states
form an eigenbasis of the fully interacting theory and have a perfectly
good interpretation as collections of asymptotically free particles with
mass and appropriate quantum numbers. (In most of the interesting
physical cases such a basis exists.) We write the expansion as
\begin{equation}
\vert0\rangle^{0}\rightsquigarrow\vert e\rangle=\vert0\rangle^{1}+\sum_{\left\{ 
a_{n}\right\} }\sum_{\left\{ p_{n}\right\} }K_{\left\{ p_{n}\right\} }^{\left\{ 
a_{n}\right\} }\vert\left\{ p_{n}\right\} \rangle_{\left\{ a_{n}\right\} 
}^{1}+\ldots,\label{eq:exp}
\end{equation}
where the eigenstates contain the stable (quasi-)particle
excitations of species $a_{n}$ and momentum $p_{n}$. Since we consider a non 
adiabatic, finite-time process the amplitudes $K_{\left\{ p_{n}\right\} 
}^{\left\{ a_{n}\right\} }$ in general will be nonzero, however, we note that for 
the multi particle states to acquire an appreciable weight the inverse time 
scale of the protocol should be much larger than the gap, $1/T\gg m$.

Now consider the pdf of the work done on the system during the protocol
defined as
\begin{equation}
P(W)=\sum_{\text{eigenstates }\vert\Psi\rangle\text{ of 
}H_1}\delta(W-E_{\Psi}+E_{gs,0})\vert\langle\Psi\vert 
e\rangle\vert^{2},\label{eq:PW}
\end{equation}
signifying two projective energy measurements before and after the
protocol and summing over all the possible transitions weighted by the
respective overlaps. Supposing a translationally invariant initial
state and time-evolving Hamiltonian the one-particle part can only
consist of zero-momentum particles responsible for Dirac deltas in
the pdf and the low energy behavior of the continuum part is dictated
by the two-particle creation amplitudes $\langle p_1 p_2\vert e\rangle= 
K(p_1)\delta(p_1+p_2)$
relative to the particles with lowest mass $m$ (only states with zero total 
momenta
are allowed because of translation invariance). 

In Ref. \cite{PalmaiSot}  for an integrable quantum
field theory in the quench limit it was observed
that if there are no particle multiplets the continuum part starts
as
\begin{equation}
P(W\gtrapprox2m)\sim\left|K(\sqrt{W^{2}-4m^{2}})\right|^{2}(W-2m)^{-1/2}\label{eq:PWsmall}
\end{equation}
where the density of states near the threshold was supposed to go as 
$\rho(E)\sim(E-2m)^{-1/2}$. Here we observe that Eq. (\ref{eq:PWsmall}) depends 
only on the relativistic dispersion $E(p)\equiv E_{\vert 
p,-p\rangle}=2\sqrt{m^2+p^2}$ and density of states and therefore generalizes to 
finite-time protocols on arbitrary interacting relativistic quantum field 
theories. Now we use the relation 
\begin{equation}
K(p)=S(-2p)K(-p),\label{eq:KSK}
\end{equation}
with $S(p)$ being the two-particle scattering amplitude. This can be verified by
considering a state $\vert\Psi\rangle=\int_{-\infty}^{\infty}dpK(p)\vert 
p,-p\rangle$ and using the definition of the scattering amplitude $\vert 
p,-p\rangle=S(2p)\vert-p,p\rangle$ to obtain 
$\vert\Psi\rangle=\int_{-\infty}^{\infty}dpK(-p)S(-2p)\vert p,-p\rangle$ proving 
Eq. (\ref{eq:KSK}). Noting that in one dimension for any interacting theory the 
scattering amplitude has the super-universal property $S(0)=-1$,\footnote{See Ref. \cite{Sachdev} after Eq. (6.13) or consider the simple quantum
mechanical problem of potential scattering, where it can be seen by
elementary considerations that in the low-energy limit, i.e. when
the potential can be approximated by a Dirac delta, the phase shift
is always $\pi$ corresponding to $S=-1$.}, we see that the two-particle 
amplitude is odd near $p=0$. The simplest
choice realizing this would be $K(p\approx0)\sim p$ giving 
$P(W\approx2m)\sim(W-2m)^{1/2}$, which was indeed observed when quenching inside 
a
single phase in the Ising \cite{SilvaSmacch} and sinh-Gordon
models \cite{PalmaiSot}. However, one could also imagine $K(p\approx0)\sim 
p^{-1}$,
or in fact any odd power. Incidentally, the choice $p^{-1}$ yields
$P(W\approx2m)\sim(W-2m)^{-3/2}$, an edge behavior observed when quenching
through the quantum critical point in the Ising model \cite{SilvaSmacch}.

In this paper we argue that in one dimension and close to criticality (or when a 
relativistic
dispersion is expected) the exponents $1/2$ and
$-3/2$ are in fact the only natural ones in any interacting system. In the 
special case of free bosons with $S(0)=1$, a 
third exponent is seen instead, $P(W\approx2m)\sim(W-2m)^{-1/2}$,
which is confirmed by explicit calculation in Ref. \cite{SGS}.

\begin{figure}
\includegraphics[scale=0.8]{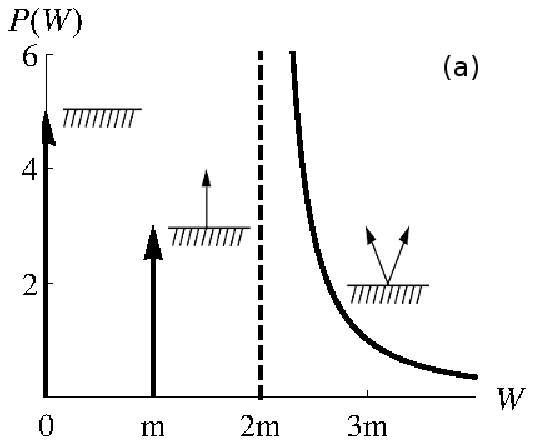}$\,\,$\includegraphics[scale=0.8]{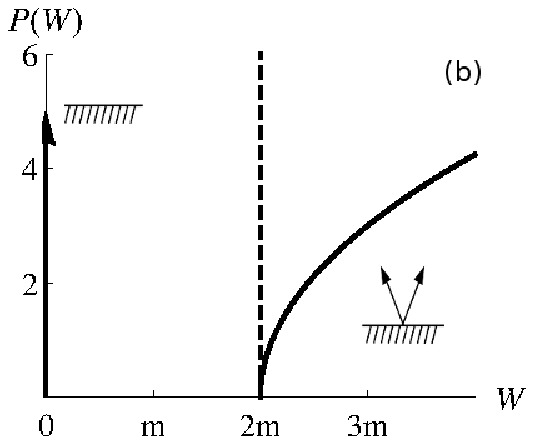}
\protect\caption{\label{fig:1}Extensivity of the initial state and a local, 
translationally
invariant, interacting Hamiltonian evolution requires a singular edge
exponent of $-3/2$ in the pdf of the work done if the protocol can
be realized by the emission of a single zero-momentum particle (a). In
the absence of the one-particle realization the edge is non-singular with
an exponent of $+1/2$ (b).}
\end{figure}

We show that the only way for an extensive quench to be realized
with a singular two-particle amplitude, e.g., $K(p\approx0)\sim p^{-1}$, is in 
the presence of a zero-momentum
one-particle excitation in the expansion (\ref{eq:exp}). Vice versa,
if there is a nonzero one-particle term in Eq. (\ref{eq:exp}), the
corresponding two-particle amplitude has a pole at $p=0$. 
To see this correspondence we note that extensivity
of free energy is expected for translationally invariant initial states
in thermodynamically large systems
because the translation operator does not change throughout the protocol.
The asymptotic expansion of
the partition function calculated in the post-protocol system in finite volume 
$L$ and
inverse temperature $R$ reads
\begin{align}
Z =1 &+a_{1}Le^{-mR}\nonumber \\
 & 
+\sum_{I}\frac{\varepsilon(p_{I})}{L\varepsilon(p_{I})+2\delta'(2p_{I})}
\left|K(p_{I})\right|^{2}e^{-2R\varepsilon(p_{I})}\nonumber \\
 & +\ldots,\label{eq:Z}
\end{align}
where the fraction in the two-particle term accounts for the difference
in the density of states in finite and infinite volumes (for details see Ref.
\cite{KormosPozsgay}, where the equivalent boundary field theoretical 
problem was considered). $\varepsilon(p)$ is the one-particle
energy at momentum $p$, $\delta(p)$ the phase shift, $S(p)=e^{i\delta(p)}$, and
$I$ labels the quantized finite volume states.
At the bottom of the spectrum the quantized momenta behave as $p_{I}\sim 
L^{-1}$,
so both $\varepsilon(p)$ and $\delta'(p)$ are finite. As shown already
in Ref. \cite{KormosPozsgay}, if the two-particle amplitude has a first-order 
pole at $p=0$, the only way for the free energy to be extensive
$F=\log Z\sim L$ is  in the presence of a nonzero one-particle contribution
and in fact the coefficient $a_{1}$ is related to the residue of
the pole of $K(p)$ \cite{BPT,KormosPozsgay}. This is because the
part of the two-particle contribution coming from the pole of $K(p)$
is superextensive of order $L^{2}$ and needs to be canceled exactly in $\log 
Z$.
\footnote{It is interesting to note that the extensivity of the non-equilibrium
protocol gives the same condition for $\vert e\rangle$ as the one
obtained in boundary field theory for sensible boundary states from 
considerations involving the
crossed channel \cite{Goshal,BPT}.} 
One can also see, that a more singular behavior of $K(p)$ at zero
cannot be canceled by the one-particle contribution, therefore we
can restrict $K(p)$ to be 
\begin{equation}
K(p\approx0)\sim p^{2k+1},\qquad k\geq-1.
\end{equation}
Considering this last equation, we expect that, without a fine tuning
in the parameters,\footnote{Fine tuning is understood in the sense that for 
different exponents to appear, $K'(0)=0$ would be required, however, the 
derivative $K'(p)$ has no simple physical meaning, and therefore this corresponds 
only to an accidental choice of protocol parameters.} the two-particle amplitude 
is linear for small momenta
unless there is a realization of the protocol with the emission of
a single particle, in which case the amplitude will have a simple pole
at $p=0$ (see Fig. \ref{fig:1}).

In the following we discuss two scenarios leading to a one-particle
contribution in the after-protocol state. In the first case the system is 
spontaneous
symmetry breaking (SSB) either before or after the protocol. The second, equally 
interesting case is when the model has a more complicated particle content, such 
as the sine-Gordon model, where bound state one-particle contributions can 
appear without crossing a critical point.

\paragraph{Spontaneous symmetry breaking.}

We take the simplest SSB system, the Ising model in transverse field
close to criticality, in the thermodynamic limit equivalent to free
massive Majorana fermions. Depending on the sign of the mass, the
system is either in the unbroken (disordered) $m>0$ or the broken symmetry phase
$m<0$ (ordered). To determine the condition for the one-particle contribution to 
appear,
we need to recall the Hilbert space structure of Majorana fermions.

The Hilbert space can be divided into two sectors, with two ground
states, according to either adopting a periodic (Neveu-Schwartz, NS)
or anti-periodic boundary condition (Ramond, R). The excitations are free
fermions and in finite volume the boundary conditions require that 
the zero-momentum excited states have even fermion numbers
relative to the ground state. In the broken phase and in the thermodynamic
limit the energies of the two ground states become degenerate, and in fact the 
two infinite volume ground states are the superpositions \cite{Weinberg}
\begin{eqnarray}
\left|\uparrow\right\rangle  & = & 
\frac{1}{\sqrt{2}}\left(\left|\text{NS}\right\rangle 
+\left|\text{R}\right\rangle \right)\nonumber \\
\left|\downarrow\right\rangle  & = & 
\frac{1}{\sqrt{2}}\left(\left|\text{NS}\right\rangle 
-\left|\text{R}\right\rangle \right)
\end{eqnarray}
The excitations over these states are kinks interpolating between the two vacua, 
i.e., moving domain walls. In the
disordered, unbroken phase the R ground state acquires a mass 
relative to the NS ground state and becomes a one-particle state, 
so the zero-momentum R
sector can be interpreted as a collection of states containing an
odd number of particles. The vacuum is the NS vacuum and the excitations
are fermions corresponding to spin waves. Now, the states from different
sectors have no overlaps with each other because they have different
topological properties, so if the initial state contains one sector, that
sector will survive any protocol.

There is an important difference between arriving
in the same or arriving in a different phase as the initial one. Let
us first take the case of starting and ending the protocol in the disordered
phase. In this case the initial state is the NS vacuum and there is
no overlap between NS and R states, so we have no one-particle contribution
in the expansion (\ref{eq:exp}). Contrary, if we start from one of
the ordered ground states and arrive in the disordered phase, because
of the presence of the R sector, initially we do expect a one-particle
contribution. The remaining cases can be obtained by the Kramers-Wannier (KW)
duality and using the fact that the work statistics has to be
identical to that of the dual protocol (the operator corresponding
to the work is invariant under the KW duality).

\begin{figure}
\begin{centering}
\includegraphics[scale=0.4]{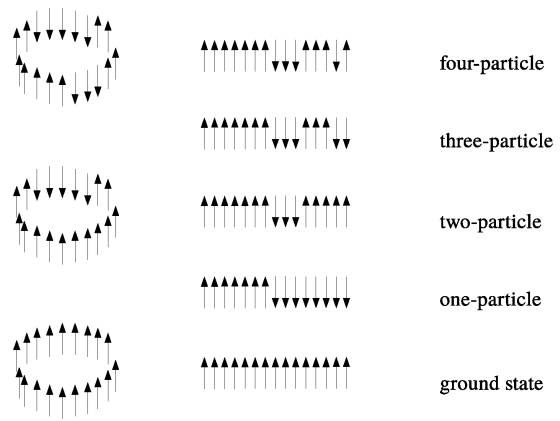}
\par\end{centering}

\protect\caption{\label{fig:2}Zero-momentum excitations in the finite and 
infinite volume Ising model in the ordered phase.}
\end{figure}

In summary, we obtained that when a protocol begins and ends in different
phases, the amplitude $K(p)$ has a pole, while if no phase boundary
is crossed it remains linear. This is in fact the correct result as
calculated in Refs. \cite{Essler,SilvaSmacch}. But contrary to the
explicit calculations available (e.g., Refs.
\cite{Essler,SilvaSmacch,SGS,SilvaDicke,PalmaiSot}),
our considerations here depended only on the structure of the Hilbert
spaces before and after the protocol, and therefore we expect them
to generalize to other SSB situations, e.g., to protocols between phases
of the three-state Potts or parafermionic models in the following way.
For discrete symmetry breaking $G\to H\subset G$, we can partition the Hilbert 
space according to the representations of $G/H$ in both the symmetric and broken 
phases, but, importantly, in the broken phase, the lowest lying states in all the 
sectors are degenerate and in infinite volume the physical vacua are linear 
combinations of these, while in the symmetric phase there is only one vacuum and 
the other sectors will contain one-particle states. Since local operators 
(relative to the Hamiltonian) have a zero matrix element between the different 
sectors, for a protocol starting in the broken phase and ending in the symmetric, 
we will in general have one-particle excitations in the expansion 
(\ref{eq:exp}).

To conclude this section we comment on the effect of finite volume on the Ising 
example. Consider the disordered-to-ordered quench when only the NS sector is 
involved. With a periodic boundary condition (PBC) there is no one-particle state in the 
broken phase in finite volume, however, in infinite volume such excitations do 
exist (see Fig. \ref{fig:2}). On the other hand, the cases of large finite and 
infinite volumes should not be qualitatively different, and indeed an explicit 
calculation of the two-particle amplitude \cite{Essler,SilvaSmacch} shows an 
infrared pole in $K(p)$ independent of the volume. Careful examination of the 
calculation for the work pdf from the exact two-particle amplitude (available 
through techniques developed for the boundary thermodynamic Bethe ansatz 
\cite{LeClair,BPT,KormosPozsgay}) shows that a finite-volume infrared 
regularization in the Ising model only allows for the appearance of the one-particle 
Dirac delta when the volume goes all the way to infinity, in accordance with the 
available excitations. These observations show that our thermodynamic argument 
connecting the one-particle contribution and the pole only works in infinite 
volume. Indeed, the finite volume vacuum $\left|\text{NS}\right\rangle $ for 
$m<0$ does not satisfy clustering and therefore an extensive free energy is not 
expected at all.

\paragraph{Bound states.}

One-particle contributions in the expansion (\ref{eq:exp}) can also
arise in models with more complicated spectra: When the post-protocol
Hamiltonian supports bound states their appearance is not forbidden
by translation and parity invariance (which was crucial in Refs. \cite{FM,SFM,STM}
to establish the structure of the after protocol state), and we expect
that generally they appear in protocols performed on such models.

To support this idea, we made numerical calculations on the sine-Gordon
model with PBCs in small volume using the truncated conformal space
approach \cite{Zamo,TakacsTCSA}. We found that, both when quenching
between the repulsive (no bound states) and the attractive (bound
states present) regimes and when quenching inside the attractive regime, there
are finite one-particle contributions in the expansion (\ref{eq:exp}) 
\cite{PTunp}.
Our predictions seem to be supported by the numerical results of Ref.
\cite{AS}, where DPTs were observed without crossing a phase
boundary for quenches in the $XXZ$ model with staggered magnetic fields in the 
parameter regime, where the low-energy reduction is the sine-Gordon model.

\paragraph{Implications for the dynamics.}

We propose that for global quenches the remarkable universality of the edge 
exponent in the work PDF can be detected in the large time behavior of the LE 
and based on whether or not a one-particle realization is allowed one can 
predict if a DPT will be encountered during time evolution. LE is defined by 
$L(t)=\vert\mathcal{L}(t)\vert^2=\vert\langle\Psi_0\vert 
e^{iH_0t}e^{-iH_1t}\vert\Psi_0\rangle\vert^2=\vert\int_{-\infty}^{\infty}dWe^{-iWt}P(W)\vert^2$ and it is connected to the work PDF 
by a Fourier transform.
To every new channel for increasing $W$
corresponds an edge with some exponent $\alpha_{nj}$ ($n$ being
the number of particles emitted in the new channel and $j$ labels
the particle species), so the long-time behavior of the Loschmidt
amplitude reads
\begin{align}
\mathcal{L}(t) = 
1+\sum_{j}b_{1j}e^{im_{j}t}&+\sum_{j}b_{2j}e^{2im_{j}t}t^{-1-\alpha_{2j}}
\nonumber \\
& +\text{higher particle terms},
\end{align}
where the first term comes from the vacuum, the second from one-particle,
and the third from two-particle contributions. Compared to the two-particle 
terms, the higher particle contributions are less singular, therefore these 
should be invisible in the
long-time limit.

For the bosonic $\alpha=-1/2$, we get $L(t)-L(\infty)\sim t^{-1/2}$,
and for the interacting $\alpha=1/2$, $L(t)-L(\infty)\sim t^{-3/2}$.
Interestingly, when there is a one-particle contribution to a given
species $j$, we would get $L(t)-L(\infty)\sim t^{1/2}$, which is
nonphysical and apparently signals that the low-energy degrees of
freedom cannot capture the long-time behavior of the LE,
and we expect nonanalytic behavior during the time evolution, or, by definition, a
dynamical phase transition.

While this is an intriguing observation, we do not suggest a one-to-one 
correspondence between
one-particle contributions in the expansion of the initial state and
DPTs. In Ref. \cite{Dora}, it was found that in
the $XY$ model it is possible to have DPTs without a singular $K(p)$
two-particle amplitude. Instead, their results also show that whenever
there is a singularity in the amplitude, there are also DPTs in the LE, supporting 
the physical relevance of the one-particle channel.

\paragraph{Conclusions.}

We proposed that the lowest edge exponents in the probability density
function of the work done during a non-equilibrium protocol correspond
to the realization of the protocol by emitting two particles and are extremely 
robust to perturbations
in gapped one-dimensional systems. In fact, in the presence of interactions,
there are only two possibilities depending on whether the protocol
can also be realized by emitting only one particle or this is forbidden. We 
discussed
two cases where such a one-particle process is allowed: when the protocol
begins and ends in different phases of a SSB model and when there
are bound states in the particle spectrum. We also proposed that if
the one-particle realization is allowed, the time evolution of the
Loschmidt echo shall exhibit a dynamical phase transition.

\begin{acknowledgments}
I thank S. Sotiriadis, G. Takacs, and M. Kormos for many useful discussions
and for reading the manuscript. I thank B. Pozsgay for clarifying some aspects
of the Ising spin chain and M. Lencses for discussions on the Loschmidt echo. 
Financial support of Hungarian Academy of Sciences, both through Grant No. 
LP2012-50/2013 and a postdoctoral fellowship, is acknowledged.
\end{acknowledgments}

\bibliographystyle{unsrtnat}

\begin{thebibliography}{KonikEssler}
\bibitem{coldat} M. Greiner, O. Mandel, T. W. Hansch, I. Bloch, 
Nature (London) 419, 51-54 (2002); T. Kinoshita, T. Wenger, D. S. Weiss, 
\emph{ibid.} 440, 900 (2006); I. Bloch, J. Dalibard and W. Zwerger, Rev. Mod. Phys. 
80, 885 (2008);
S. Trotzky Y.-A. Chen, A. Flesch, I. P. McCulloch, 
U. Schollwock, J. Eisert, I. Bloch, Nature Phys. 8, 325 (2012);
M. Gring, M. Kuhnert, T. Langen, T. Kitagawa, B. Rauer, M. Schreitl, 
I. Mazets, D. Adu Smith, E. Demler and J. Schmiedmayer, Science 337, 
1318 (2012); T. Fukuhara, P. Schauss, M. Endres, S. Hild, M. Cheneau, I. Bloch 
and C. Gross, Nature (London) 502, 76 (2013)

\bibitem{Heyl}M. Heyl, A. Polkovnikov, S. Kehrein, Phys Rev.
Lett. 110, 135704 (2013)

\bibitem{Heyl2}M. Heyl, Phys. Rev. Lett. 113, 205701 (2014)

\bibitem{SilvaSmacch}P. Smacchia, A. Silva, Phys. Rev. E 88,
042109 (2013)

\bibitem{AS}F. Andraschko, J. Sirker, Phys. Rev. B 89, 125120
(2014)

\bibitem{Dora}S. Vajna, B. Dora, Phys. Rev. B 89, 161105(R) (2014)

\bibitem{newrefs} C. Karrasch and D. Schuricht, Phys. Rev. B 87, 195104 (2013); 
M. Fagotti, arXiv:1308.0277; 
E. Canovi, P. Werner, and M. Eckstein, Phys. Rev. Lett. 113, 265702 (2014); 
J. M. Hickey, S. Genway, and J. P. Garrahan, Phys. Rev. B 89, 054301 (2014); 
J. N. Kriel, C. Karrasch, and S. Kehrein, \emph{ibid.} 90, 125106 (2014); 
M. Schmitt and S. Kehrein, \emph{ibid.} 92, 075114 (2015); 
A. J. A. James and R. M. Konik, \emph{ibid.} 92, 161111 (2015); 
S. Vajna and B. Dora, \emph{ibid.} 91, 155127 (2015) 

\bibitem{Silva2008}A. Silva, Phys. Rev. Lett. 101, 120603
(2008)

\bibitem{Jarz}C. Jarzynski, Phys. Rev. Lett. 78, 2690(1997); 
G. E. Crooks, Phys. Rev. E 60, 2721 (1999); 
H. Tasaki, arXiv:cond-mat/0009244; 
M. Campisi, P Hanggi, P. Talkner, Rev. Mod. Phys. 83, 771 (2011)

\bibitem{Talkner}P. Talkner, E. Lutz, P. Hanggi, Phys. Rev.
E 75, 050102(R) (2007)

\bibitem{Roncaglia}A. J. Roncaglia, F. Cerisola, J. P.
Paz, Phys Rev. Lett. 113, 250601 (2014)

\bibitem{PalmaiSot}T. Palmai, S. Sotiriadis, Phys. Rev. E 90,
052102 (2014)


\bibitem{GS}A. Gambassi, A. Silva, Phys. Rev. Lett. 109, 250602
(2012)

\bibitem{KonikEssler}F. H. L. Essler, R. M. Konik, \emph{From Fields to Strings: Circumnavigating theoretical Physics}, edited by
M. Shifman, A. Vainshtein, J. Wheater (World Scientific, Singapore, 2005), Vol. 1, p. 684

\bibitem{Sachdev}S. Sachdev, \emph{Quantum Phase Transitions},
1st ed. (Cambridge University Press, Cambridge U.K., 1999)

\bibitem{SGS}S. Sotiriadis, A. Gambassi, A. Silva, Phys. Rev.
E 87, 052129 (2013)

\bibitem{KormosPozsgay}M. Kormos, B. Pozsgay, JHEP 2010:112 (2010)

\bibitem{BPT}Z. Bajnok, L. Palla, G. Takacs, Nucl. Phys. B 716,
519 (2005)

\bibitem{Goshal}S. Goshal, A. Zamolodchikov, Int. J. Mod. Phys.
A 09, 3841 (1994)

\bibitem{Weinberg} S. Weinberg, \emph{The Quantum Theory of Fields}, Vol. 2 (Cambridge University Press, Cambridge U.K., 1996)


\bibitem{Essler}P. Calabrese, F. H. L. Essler, M. Fagotti,
Phys. Rev. Lett. 106, 227203 (2011); J. Stat. Mech. 2012, P07016


\bibitem{SilvaDicke}F. N. C. Paraan, A. Silva, Phys.
Rev. E 80, 061130 (2009)

\bibitem{LeClair}A. LeClair, G. Mussardo, H. Saleur, S. Skorik, Nucl. Phys. B 
453, 581 (1995)


\bibitem{SFM}S. Sotiriadis, D. Fioretto, G. Mussardo, J. Stat.
Mech. 2012 P02017

\bibitem{FM}D. Fioretto, G. Mussardo, New J. Phys. 12 055015
(2010)

\bibitem{STM}S. Sotiriadis, G. Takacs, G. Mussardo, Phys. Lett.
B 734, 52 (2014)


\bibitem{Zamo}V. P. Yurov and Al. B. Zamolodchikov, Int. J. Mod.
Phys. A 05, 3221 (1990)

\bibitem{TakacsTCSA}G. Feverati, F. Ravanini, G. Takacs,
Phys. Lett. B 430, 264 (1998)

\bibitem{PTunp}T. Palmai (unpublished)

\end{thebibliography}

\end{document}